# Computing Educator Attitudes about Motivation

Amber Settle
DePaul University
243 S. Wabash Avenue
Chicago, IL 60604
asettle@cdm.depaul.edu

Brian Sedlak
DePaul University
243 S. Wabash Avenue
Chicago, IL 60604
briansedlak9@gmail.com

## ABSTRACT
While motivation is of great interest to computing educators, relatively little work has been done on understanding faculty attitudes toward student motivation. Two previous qualitative studies of instructor attitudes found results identical to those from other disciplines, but neither study considered whether instructors perceive student motivation to be more important in certain computing classes. In this work we present quantitative results about the perceived importance of student motivation in computing courses on the part of computing educators. Our survey results show that while a majority of respondents believe student motivation is necessary in all computing courses, the structure and audience in certain computing classes elevate the importance of student motivation.

## Categories and Subject Descriptors
K.3.2 [**Computer and Information Science Education**]: Computer science education

## General Terms
Measurement

## Keywords
Instructor attitudes, student motivation

## 1. INTRODUCTION
Student motivation is a topic of great interest to computing educators, since instructors would like to see students reach their full potential and typically work from the understanding that students who are highly motivated will learn more [2]. This quite naturally means that a majority of the relevant articles in the computing education literature focus on the impact of activities and curricula on student motivation.

Evidence for the emphasis on approaches aimed at students can be seen in the results of two working groups on student motivation [1, 2]. The focus of the first working group was on motivating top students and particularly on motivating experienced students in introductory courses since that is the place in the curriculum most likely to have large gaps in ability levels between students [1]. The working group reviewed the literature and did a survey of instructors, classifying the approaches taken in introductory programming courses to help motivate experienced and talented students. The second working group also focused on programming courses and approaches but considered the student body more broadly and classified motivational approaches used by faculty as well as student perceptions of those approaches [2].

As may be surmised from the focus of the working groups on student motivation, it is not the case that student motivation is seen as uniformly important in all areas of computing. For example, despite the wide attention paid to motivating programming students, not all programming classes or topics receive equal attention with respect to motivation [8]. Recursion in particular is an area where motivation is underrepresented [7].

Understanding the connection between computing faculty attitudes and student motivation is also an area where relatively little research has been done. Much more work on instructor attitudes has been done in other disciplines. In the literature on student reading and achievement, a study found a significant gap between instructor perceptions of student motivation and actual levels of student motivation [6]. Additionally, a study of the application of natural language processing techniques to understanding student-tutor dialogs while learning physics found that improvements in motivation during tutoring were tied to dialog cohesion [9]. However, there are at least two studies in the computing education literature that have considered instructors' viewpoints of student learning, success, and motivation.

One study of instructor attitudes did a phenomenographic investigation of instructor perception of factors contributing to student success, finding that there were five categories of factors: subject, intrinsic, previous experience, attitude and behavior, and developmental [4]. Motivation in particular was mentioned in the attitude and behavior category, although specifics about what the instructors believed regarding motivation were not reported in the article. The authors also mapped the categories to the existing computing education literature on student success and found both studies that supported instructors' beliefs and gaps between instructors' beliefs and existing research [4].

A second phenomenographic study of instructors' beliefs directly considered instructor attitudes about student motivation [5]. The researchers identified four understandings of motivation: transfer (motivation as coming from the instructor), shaping (motivation as something to be developed within the student), travelling (motivation as something determined by the journey's path), and growing (motivation as something to be cultivated within and by the student) [5, pg. 103]. In the transfer understanding, the instructor's engagement directly affects student engagement. In shaping, motivation is something developed by students under the influence of the instructor. In the travelling understanding, it is the details in the material that help with student understand, such as the examples used in class. And in the growing view, student motivation begins in the shaping category which later shifts as students begin to motivate themselves [5]. The authors concluded that their work supported previous research on categories of teaching, albeit work that was done outside of computing [3].

Although it has not been widely studied in the computing education literature, instructor attitudes about student motivation are important. Instructors who do not believe that student motivation is worthwhile in a given context will not implement even well-known activities for improving student engagement. In this work we address the question of instructor attitudes and background in relation to student motivation. In particular, we are interested in answering the following research questions:

1. Do computing faculty believe that student motivation is equally important in all subjects in computing?
2. Do computing faculty believe that student motivation is equally important at all levels (e.g. introductory and advanced) in the computing curriculum?
3. Do computing faculty believe that students need to be equally motivated by all courses in their major?
4. Does the graduate training computing faculty received have an impact on faculty beliefs regarding student motivation in computing courses?
5. Does the academic position, the number of years faculty have been teaching computing courses, or gender impact faculty beliefs regarding student motivation in computing courses?

To measure computing educator attitudes about the importance of student motivation in various courses and subjects in the computing curriculum, we developed a survey which was made available to members of two computing education communities during May 2014. We analyzed the responses to understand what faculty believe about student motivation in computing courses as well as to determine if there are any connections between gender, instructor training, experience, and attitudes about motivation.

## 2. METHODOLOGY
In this section we describe the survey and participant recruitment.

### 2.1 The Survey
To understand the demographics of the participants in the study and to measure computing educators' attitudes about student motivation, we developed a brief survey. The first four questions gather basic demographic information and the remaining seven questions provide information about attitudes toward motivation in various places in the computing curriculum.

1. Please indicate your sex
2. I am a: choose one of (tenured faculty member, tenure-track faculty member, non-tenure-track, fulltime faculty member, adjunct faculty member, emeritus faculty member, K-12 teacher, graduate teaching assistant, other)
3. My graduate (Masters, Ph.D.) study was/is in: choose one of (computer science: education, computer science: hardware, computer science: networks, computer science: software engineering, computer science: systems, computer science: theory, computer science: other, human-computer interaction, information systems, information technology, other, I did not do graduate work)
4. I have been teaching computing for: choose one of (0-4 years, 5-10 years, 11+ years, I do not teach computing courses)
5. Please rank the importance of student motivation with respect to learning in the following types of courses with 1 = motivation is most important, 2 = motivation is second most important, etc.
    a. Courses for non-computing majors
    b. Introductory courses for computing majors
    c. Advanced, required courses for computing majors
    d. Advanced, elective courses for computing majors
6. Please indicate why you ranked student motivation as indicated in the previous question.
7. Please rank the importance of student motivation with respect to learning in the following subjects, with 1 = motivation is most important, 2 = motivation is second most important, etc.
    a. Programming courses
    b. Networking courses
    c. Theory courses
    d. Systems courses
8. Please indicate why you ranked student motivation as indicated in the previous question.
9. Please indicate your agreement with this statement: A student does not have to have equal motivation for each computing course he/she takes: strongly disagree / disagree / neither agree nor disagree / agree / strongly agree
10. Please indicate your agreement with this statement: Overall student motivation for the program of study matters more than the motivation for any single course: strongly disagree / disagree / neither agree nor disagree / agree / strongly agree
11. Students can find some computing subjects more motivating than others. Please indicate the computing subject(s) you believe are the most motivating for students.

### 2.2 Participant Recruitment
Participants were recruited via emails sent to the mailing lists of the ACM Special Interest Groups for Computer Science Education (SIGCSE) and Information Technology Education (SIGITE). After approval of the protocol by the DePaul University Institutional Review Board a request to complete the survey was sent to both the SIGCSE-members and SIGITE-members mailing list. At the time of the study the SIGCSE-members mailing list had 1220 subscribers and the SIGITE-members mailing list had 867 subscribers so that a total of 2087 people were contacted.

Participants were directed to an online version of the survey developed using the Qualtrics online survey tool. Only participants who agreed to the letter of consent on the first page of the survey were able to complete it. The letter of consent ensured that the participants were at least 18 years of age and agreed to be a part of the study.

## 3. RESULTS
There were 108 complete responses to the survey, for a response rate of 5.17%.

### 3.1 Demographics
There were 80 responses from men (74%), 27 responses from women (25%), and 1 response from a person who preferred not to specify gender (1%).

Table 1 provides information about the positions held by respondents. The respondents who indicated other included a Masters student who is an adjunct in another division, an IT professional, and a faculty member at a non-tenure campus.

## Table 1. Positions held by respondents

| Position | N | % |
|---|---|---|
| Tenured faculty | 64 | 59% |
| Tenure-track faculty | 8 | 7% |
| Non-tenure track, fulltime faculty | 17 | 16% |
| Adjunct faculty | 10 | 9% |
| Emeritus faculty | 3 | 3% |
| K-12 teacher | 0 | 0% |
| Graduate teaching assistant | 3 | 3% |
| Other | 3 | 3% |

A majority of respondents had done graduate (Masters or Ph.D.) work in a subfield of computer science. Table 2 provides information about the graduate training of respondents.

## Table 2. Graduate training of respondents

| Graduate work | N | % |
|---|---|---|
| CS: education | 5 | 5% |
| CS: hardware | 1 | 1% |
| CS: networks | 4 | 4% |
| CS: software engineering | 6 | 5% |
| CS: systems | 12 | 11% |
| CS: theory | 17 | 16% |
| CS: AI | 8 | 7% |
| CS: other | 17 | 16% |
| HCI | 4 | 4% |
| IS | 8 | 7% |
| IT | 5 | 5% |
| Other | 20 | 18% |
| I did not do graduate work | 1 | 1% |

The "computer science: other" category included responses for algorithms, algorithms and machine learning, computational science, educational technology, high-performance computing, intelligent systems, programming languages (2), symbolic/algebraic computation, systems analysis and design. There were sufficient responses in the other category for artificial intelligence that those have been separated in the table above. The other category included astrophysics, business education, education (3), educational psychology, information assurance, instructional technology, industrial technology, information technology education, mathematics (6), MBA, organizational management (2), and systems technology.

A majority of respondents had been teaching computing for at least 11 years. Table 3 summarizes the respondents' teaching experience.

## Table 3. Years teaching computing

| Range | N | % |
|---|---|---|
| 0-4 years | 4 | 4% |
| 5-10 years | 16 | 15% |
| 11+ years | 87 | 80% |
| I do not teach computing courses | 1 | 1% |

### 3.2 Importance of Motivation

Table 4 summarizes the responses of faculty when asked about the importance of student motivation in various levels of computing courses, including introductory courses for majors and non-majors along with advanced courses for majors. A 1 indicated that motivation was most important, a 2 that motivation was second most important, etc. Respondents were allowed to rank items equally.

## Table 4. Ranking of motivation by course level and audience

| Type | 1st | 2nd | 3rd | 4th | Mean |
|---|---|---|---|---|---|
| Courses for non-computing majors | 51 | 29 | 12 | 16 | 1.94 |
| Introductory courses for computing majors | 71 | 27 | 9 | 1 | 1.44 |
| Advanced, required courses for computing majors | 48 | 34 | 18 | 8 | 1.87 |
| Advanced, elective courses for computing majors | 46 | 25 | 22 | 15 | 2.06 |

Table 5 summarizes the responses of faculty when asked about the importance of student motivation in subjects within the computing curriculum including programming, networking, theory, and systems courses. A 1 indicated that motivation was most important, a 2 that motivation was second most important, etc. Respondents were allowed to rank items equally.

## Table 5. Ranking of motivation by course subject

| Subject | 1st | 2nd | 3rd | 4th | Mean |
|---|---|---|---|---|---|
| Networking | 37 | 40 | 18 | 13 | 2.06 |
| Programming | 75 | 20 | 11 | 2 | 1.44 |
| Systems | 46 | 37 | 17 | 8 | 1.88 |
| Theory | 69 | 25 | 8 | 6 | 1.55 |

Table 6 summarizes responses asking for agreement about the statement: "A student does not have to have equal motivation for each computing course he/she takes."

The overwhelming consensus was that equal motivation for each course is not necessary, with 77% of responses either agreeing or strongly agreeing with the statement.

Table 7 summarizes responses asking for agreement about the statement: "Overall student motivation for the program of study matters more than the motivation for any single course."

Again a strong majority (68%) agreed or strongly agreed that overall motivation was more important than a single course.

**Table 6. Students need equal motivation for each course**

| Response | Count | Percentage |
|---|---|---|
| Strongly agree | 32 | 30% |
| Agree | 51 | 47% |
| Neither agree nor disagree | 8 | 7% |
| Disagree | 9 | 8% |
| Strongly disagree | 8 | 7% |

**Table 7. Overall motivation matters more**

| Response | Count | Percentage |
|---|---|---|
| Strongly agree | 33 | 31% |
| Agree | 40 | 37% |
| Neither agree nor disagree | 16 | 15% |
| Disagree | 14 | 13% |
| Strongly disagree | 5 | 5% |

## 3.3 Open-ended Responses

There were three open-ended questions on the survey. Two of them followed the questions asking respondents to rank the importance of motivation in various levels of computing classes (e.g. non-majors, introductory, etc.) and in various types of computing classes (e.g. systems, theory, etc.). The final question on the survey asked about courses in which students are particularly motivated. In this section we discuss results from the first two open-ended questions. The analysis of the third question is not presented here to do space limitations.

All of the responses for the first open-ended question were read by both authors, who independently produced categorizations of the responses. Disagreements were discussed until a final set of categories was produced. The second author then categorized each response, and the first author reviewed those categorizations. Disagreements were discussed until there was consensus.

When asked why they had ranked the importance of student motivation for various types of computing courses (non-major, introductory courses for the major, required advanced courses, and elective advanced courses), 98 of the participants responded. The following were the statements distilled from the responses by the authors:

1. Motivation is important and a prerequisite for all learning situations.
2. The structure of the class (e.g. type of topic, type of activity, amount of context, etc.) impacts the importance of motivation.
3. Motivation is important for courses with significant obstacles (e.g. challenge, lack of context, inherent frustration, etc.).
4. Motivation impacts persistence (e.g. in a major/on a problem/to move past obstacles/to overcome a lack of confidence).
5. Other student characteristics (e.g. ability, willingness to learn, maturity, work ethic, etc.) are more important than motivation in some situations (e.g. types of courses, types of students).
6. There is a relationship between instructor excitement/motivation and student motivation.
7. One of the roles of the instructor is to motivate students.
8. Student motivation has multiple sources (e.g. subject interest, perceived subject value, grades, etc.).
9. Computing students should be/are inherently motivated by computing courses.
10. Students in non-major computing courses are less likely to be self-motivated.
11. Motivation is important in courses that are used to attract students (e.g. non-majors, introductory courses).
12. Students in introductory courses are more likely to be unmotivated.
13. Students in advanced courses are more likely to be motivated.
14. Self-selected classes (e.g. non-majors, advanced electives, etc.) require less motivation.
15. A lack of motivation is detrimental in elective courses.
16. Required classes (e.g. non-majors, required computing classes, etc.) require less motivation.
17. Required classes (e.g. non-majors, required computing classes, etc.) require less motivation.

The table below summarizes the frequency with which the 98 respondents agreed with the statements above. A single response could, and often did, generate multiple statements, so the percentages sum to more than 100. The statements are listed with those receiving the highest percentage first.

**Table 8. Classification of open-ended responses to Q6**

| Statement | Percentage | Statement | Percentage |
|---|---|---|---|
| 1 | 51% | 5 | 11% |
| 2 | 49% | 17 | 11% |
| 11 | 27% | 14 | 10% |
| 4 | 24% | 8 | 9% |
| 3 | 23% | 15 | 7% |
| 7 | 17% | 12 | 6% |
| 13 | 15% | 6 | 3% |
| 9 | 12% | 16 | 1% |
| 10 | 12% | | |

When asked why they had ranked the importance of student motivation in various computing subjects (networking, programming, systems, theory), 92 of the participants responded. As with the previous open-ended question, the authors distilled statements from the responses. The first five statements were identical to the above question. Below are the latter nine statements for Q8:

6. One of the roles of the instructor is to motivate students.
7. Student motivation has multiple sources (e.g. subject interest, perceived subject value, grades, etc.).
8. Computing students should be/are inherently motivated by computing courses.

9. Students in courses that are less directly related to their field of study are less likely to be motivated.
10. Students in courses that are directly related to their interests or future career are more likely to be motivated.
11. Students are more motivated by the material in programming classes.
12. Students are more motivated by the material in networking courses.
13. Motivation is more important for students in theory courses.
14. Motivation is more important for students in systems courses.

The table below summarizes the frequency with which the 92 respondents agreed with the statements above. As with the previous question the percentages sum to more than 100. The statements are listed with those receiving the highest percentage first.

**Table 9. Classification of open-ended responses to Q8**

| Statement | Percentage | Statement | Percentage |
|---|---|---|---|
| 2 | 39% | 14 | 10% |
| 1 | 27% | 7 | 9% |
| 13 | 27% | 10 | 7% |
| 11 | 13% | 4 | 5% |
| 3 | 12% | 6 | 4% |
| 12 | 12% | 9 | 3% |
| 5 | 11% | 8 | 1% |

### 3.4 Differences in Subgroups

In analyzing the data from the surveys we considered subpopulations of respondents to determine if there were any statistically significant differences in their responses to any of the ranking or Likert-scale questions. The groups we considered were male versus female respondents, respondents in various academic positions, respondents with graduate training in particular areas, and seniority of respondents.

For the gender differences, the responses were analyzed using a t-test for independent samples, and for the remaining subpopulations the responses were analyzed using a one-way ANOVA test. The null hypothesis in each case was that the subpopulations did not differ with respect to their ranking or responses for the associated question. The analysis was done using SPSS.

There were no significant differences between any of the subpopulations on the ranking questions for course level or course type or on the final two Likert-scale questions.

### 4. DISCUSSION

The majority of survey respondents was male (74%), tenured (59%), did their graduate work in an area of computer science (65%), and had been teaching for at least 11 years (80%).

### 4.1 The Importance of Motivation

In all levels of courses a majority of faculty rated student motivation as being either most important or second most important. There was the strongest agreement for introductory courses for majors, with 71 (66%) of respondents indicating that motivation was most important and 98 (91%) indicating that motivation was either first or second most important in those courses. The type of course with the next greatest support for the importance of motivation was courses for non-computing majors. Fifty-one respondents (47%) indicated motivation was most important and 80 (74%) indicated motivation was either first or second most important in those courses. The importance of motivation in advanced courses for majors was rated less highly. Forty-eight (44%) rated motivation as most important and 82 (76%) rated it as first or second most important in advanced, required courses for majors. 46 (43%) rated motivation as most important and 71 (66%) rated it as first or second most important in advanced, elective courses for majors.

In all types of courses a majority of faculty rated motivation as being either important or second most important. There was the strongest agreement for programming courses, with 75 (69%) of respondents indicating that motivation was most important and 95 (88%) indicating that motivation was either first or second most important in those courses. A close second in terms of the importance of motivation was theory courses. 69 respondents (64%) rated motivation most important in those courses, and 94 (87%) indicated that motivation was either first or second most important. For systems classes 46 respondents (43%) indicated motivation was most important in those courses and 83 (77%) indicated that motivation was most or second most important. The weakest support for the importance of motivation was in networking courses. Only 37 (34%) rated motivation as most important and only 77 (71%) rated motivation as most or second most important in networking courses.

The results of this work do not provide evidence that gender, graduate training, academic position, or seniority has an influence on the importance computing educators place on motivation. There were no statistically significant differences found in responses between any of those subpopulations. It is possible that the lack of diversity in the data collected may have influenced the results. The areas of graduate study for the respondents were also splintered into many small subgroups, making it more difficult to obtain statistically significant results in the responses.

### 4.2 Open-Ended Questions

This work did provide insight into faculty views on motivation, particularly when the classifications of the open-ended questions are considered. A majority of respondents (Q6: 51%) indicated that motivation is important in all computing classes, with many of them objecting to the questions that asked them to rate motivation in different types of classes. These respondents believe that motivation is a prerequisite to all learning. Close to a majority of respondents (Q6: 49%, Q8: 39%) indicated that the structure of a class, including the amount of context provided and the type of computing topic, influences the importance of motivation. More than one in five respondents agreed that motivation is important in courses designed to attract students (Q6: 27%), that it has an impact on persistence (Q6: 24%), and that it is important for courses with significant obstacles (Q6: 23%). More than a quarter of respondents agreed that motivation is important for theory courses (Q8: 27%). More than one in ten respondents agreed that students are more likely to be motivated by programming classes (Q8: 13%) and by networking classes (Q8: 12%). Roughly the same proportion (Q6: 17%) believe that one of the instructor's roles is to motivate students.

Information about where motivation is not perceived to be important could also be seen in the results. Less than one in ten respondents indicated that a lack of motivation is detrimental in elective courses (Q6: 7%). Respondents also did not perceive there to be strong connections between instructor and student motivation (Q6: 3%).

### 4.3 Limitations
The data which was analyzed included responses only from computing educators subscribed to the SIGCSE and SIGITE mailing lists. These educators are more likely to be familiar with studies on motivation and student learning, which may bias their opinion regarding the importance of student motivation. A survey of all computing educators might therefore produce different results.

Several of the subpopulations, particularly the type of graduate training, were small, making statistically analysis of those groups impossible.

The survey questions were reviewed by people other than the researchers participating in the work in order to ensure clear and valid responses, but several of the open-ended comments indicated that some respondents found them to be unclear. This may have impacted the results.

## 5. ACKNOWLEDGMENTS
Our thanks to Valerie Barr, André Berthiaume, and Andrew Luxton-Rielly for their suggestions regarding the survey.

## 6. CONCLUSION
The work presented here provides more insight into faculty attitudes about motivation. Responses from computing educators drawn from SIGCSE and SIGITE mailings lists showed that the majority perceive motivation to be important in all learning situations, but particularly in courses designed to attract majors, in courses with significant obstacles, and in particular areas of computing, including theoretical courses. Some computing educators also believe that programming and networking courses are more motivating for students. This work provides quantitative results to complement existing qualitative studies on faculty attitudes toward student motivation.